\documentclass[a4paper,fleqn]{cas-dc}

\usepackage[authoryear,longnamesfirst]{natbib}
\usepackage{tabularx}
\usepackage{listings}

\def\tsc#1{\csdef{#1}{\textsc{\lowercase{#1}}\xspace}}
\tsc{WGM}
\tsc{QE}

\usepackage{xcolor}
\usepackage{ulem}
\newcommand{\add}[1]{#1}
\newcommand{\del}[1]{}

\begin{document}
\let\WriteBookmarks\relax
\def\floatpagepagefraction{1}
\def\textpagefraction{.001}

\shorttitle{Automated Photometric Pipeline for XZUT-80LCP}    
\shortauthors{Xu et al.}  

\title [mode = title]{An Automated Photometric Pipeline for the 80cm Xizang University Telescope}  



%











\author[lcr,xzu]{Xu Chao}\ead{373849362@qq.com}

\author[naoc]{Zheng Jie}\cormark[1]\ead{jiezheng@nao.cas.cn}
\credit{Conceptualization, Software, Supervision, Writing – review \& editing}
\author[lcr,xzu]{Chen Tian-Lu}\cormark[1]\ead{chentl@utibet.edu.cn}
\credit{Funding acquisition, Project administration, Resources, Writing – review \& editing}

\author[lsnu1,lsnu2,lsnu3]{Jiang Lin-Qiao}
\author[lcr,xzu]{Bao Hua}
\author[lcr,xzu]{Li Ying-Gang}
\author[lcr,xzu]{SuoNan-DaJi}
\author[fxl]{Feng Xing-Lan}

\affiliation[lcr]{organization={The Key Laboratry of Cosmic Rays (Xizang University), Ministry of Education},
            city={Lhasa},
            postcode={850000}, 
            state={Xizang},
            country={China}}
            
\affiliation[xzu]{organization={Department of Physics, College of Science, Xizang University},
            city={Lhasa},
            postcode={850000}, 
            state={Xizang},
            country={China}}
            
\affiliation[naoc]{organization={National Astronomical Observatories, Chinese Academic of Sciences},
            city={Chaoyang},
            postcode={100101}, 
            state={Beijing},
            country={China}}

\affiliation[lsnu1]{organization={College of Physics and Optoelectronic Engineering, Leshan Normal University}, 
            city={Leshan},
            postcode={614000}, 
            state={Sichuan},
            country={China}}

\affiliation[lsnu2]{organization={Key Laboratory of Detection and Application of Space Effect in Southwest Sichuan}, 
            city={Leshan},
            postcode={614000}, 
            state={Sichuan},
            country={China}}            
            
\affiliation[lsnu3]{organization={Center for Applied Optics Research, Leshan Normal University}, 
            city={Leshan},
            postcode={614000}, 
            state={Sichuan},
            country={China}}      

\affiliation[fxl]{organization={The Experimental High School Attached to Beijing Normal University}, 
            city={Xicheng},
            postcode={100032}, 
            state={Beijing},
            country={China}}

\cortext[1]{Corresponding Authors: Zheng Jie, Chen Tian-Lu}

\begin{abstract}
Processing astronomical data can take up a significant amount of researchers' time. The 80cm telescope at Xizang University is currently in its trial operation phase; however, it lacks a data processing program, which makes efficient handling of the data it generates an urgent concern. To address this issue, we have developed an automatic pipeline for processing photometric data and extracting light curves using Python 3. This pipeline has several advantages, including high speed, ease of use, and modularity. The differential photometric accuracy of this pipeline is comparable to that of data processing programs used by other similar telescopes. This development effectively overcomes the limitations of manually processing data, providing efficient and reliable support for future studies of variable stars. The pipeline has already been integrated into the telescope's operational system.
\end{abstract}



\nocite{*}

\begin{keywords}
photometric data processing \sep variable star \sep light curve \sep differential photometry \sep pipeline
\end{keywords}

\maketitle


\section{Introduction}

Photometric observations of variable stars are an important area of research in Astronomy. Common targets in variables include eclipsing binaries, exoplanets, Cepheid, and RR Lyrae variable \citep{AAVSO2011}. Each observation generates a large amount of raw data. Effectively processing this data has always been a challenge for researchers.

The 80cm Xizang University Telescope (XZUT-80) is located on the roof of Building 14 at Najin Campus of Xizang University. Its coordinates are: longitude $91^\circ10.8'$E, latitude $29^\circ39'$N, and elevation approximately 3,650m. XZUT-80 is a telescope system comprising a primary telescope, an equatorial mount, a guide telescope, a finder telescope, and an imaging system. Its optical set uses the Ritchey-Chretien system. The primary mirror has a diameter of 80cm and a focal ratio of F/8. The telescope is mounted on a fork-type equatorial mount. Its terminal is equipped with a ProLine CCD camera, model 23042, manufactured by FLI and featuring electronic cooling. The camera is placed at the Cassegrain focus of the telescope. For more camera parameters, please refer to Table~\ref{ccd_parameters}. Johnson-UBVRI broadband filters are installed between the telescope and the camera. It also supports clear and none filter modes. The average atmospheric seeing at the site is approximately $1.43''$. For more information about the telescope and its site, please refer to \citet{Bao2024}. Once the observation is complete, the camera can output raw image data in the standard FITS format. The XZUT-80 has now produced research results in two fields: exoplanets \citep{SuonanDajietal2025} and eclipsing binaries \citep{Lietal2025}. However, since the telescope system is not equipped with a data processing program, observers must manually process the raw image data, which is a time-consuming process. 

\begin{table}
\caption{Main parameters of the CCD camera for the XZUT-80.}\label{ccd_parameters}
\begin{tabular*}{\tblwidth}{@{}LL@{}}
\toprule
Parameter & Value \\
\midrule
CCD pixel count & $2048 \times 2048$ \\
CCD pixel size & $15\,\mu m \times 15\,\mu m$ \\
Readout speed & 0.5 MHz, 3 MHz \\
Readout noise & $13\,e^{-}$, $22\,e^{-}$ \\
Pixel scale & $0.48''$/pixel \\
Filed of view & $16.7' \times 16.7'$ \\
\bottomrule
\end{tabular*}
\end{table}

Many large telescopes and their associated observational system are equipped with a corresponding data processing pipeline. For example, the Transiting Exoplanet Survey Satellite (TESS) uses the QuickLook Pipeline (QLP) to process TESS full-frame images (FFIs) data \citep{Huangetal2020}. The NSF-DOE Rubin Observatory has developed a data processing software stack for optical/infrared surveys, known as the LSST Data Management System \citep{Juricetal2017}. In China, the Wide Field Survey Telescope (WFST) has customised the LSST stack using the self-developed obs\_wfst package to process WFST data \citep{Caietal2025}. The SiTian project is a large-scale, time-domain survey based on a general-purpose telescope array and wide-field CMOS detectors. The Mini-SiTian Array, the SiTian project's pathfinder, has a specially designed set of high-precision data processing pipelines for wide-field CMOS detectors \citep{Xiaoetal2025}. While some of these programs are publicly available, others are integrated into telescope service systems. This program suite will provide excellent processing results for telescope data and enhance observation efficiency. In most cases, these specialised programs cannot be used directly to process data from other telescopes. 

Over the years, the astronomical community has developed various data processing software for different research purposes and is freely available. However, these efforts are often fragmented and uncoordinated. In Python, image alignment - a crucial step in photometric data processing - involves multiple software packages, each of which uses significantly different implementation mechanisms. In most cases, results processed by different software packages cannot be used directly with one another. Currently, photometric data processing still uses Image Reduction and Analysis Facility (IRAF), a collection of software widely used in scientific research \citep{Tody1986}. The software can run on all major operating systems and includes a wide range of software packages. These packages have been developed for specific research areas or observational facilities and are adequate to meet the needs of researchers. However, this software is no longer updated and maintained, and is actually quite complicated to use. In conclusion, although there are many tools related to photometric data processing, it was not straightforward to transfer them to a telescope that had just started operating. 

Taking the above into account, the author has developed an automatic data processing pipeline that processes variable star data generated by the XZUT-80 and extracts light curves. This pipeline is named the 80cm Xizang University Telescope Light Curve Pipeline (XZUT-80LCP). Its workflow follows the standard step for CCD multiband photometry data processing \citep{YangWu2012}. The paper focuses on the introduction, application and processing results of this pipeline. The author argues that observers can accurately and reasonably plan subsequent work based on the pipeline's output.

\section{Program Framework}

\subsection{Supporting Environment}

Over the past decade, the Python programming language has rapidly gained popularity in the field of astronomy, primarily due to its user-friendly nature \citep{Greenfield2011}. Many researchers have created various astronomy-related packages using Python to aid in their projects. The XZUT-80LCP is written in Python 3 and utilizes multiple Python packages.

In the source detection and photometry module, the pipeline employs the astronomical software SExtractor \citep{BertinArnouts1996} to perform necessary tasks. The image alignment module uses the qmatch package, as referenced in Article \citet{Zhengetal2024}. Furthermore, the program relies on additional packages such as Astropy \citep{AstropyCollaborationetal2022}, NumPy \citep{Charlesetal2020}, and Matplotlib \citep{Hunter2007}.

\subsection{Pipeline Flow}

\begin{figure}
    \centering
   	\includegraphics[width=0.4\textwidth]{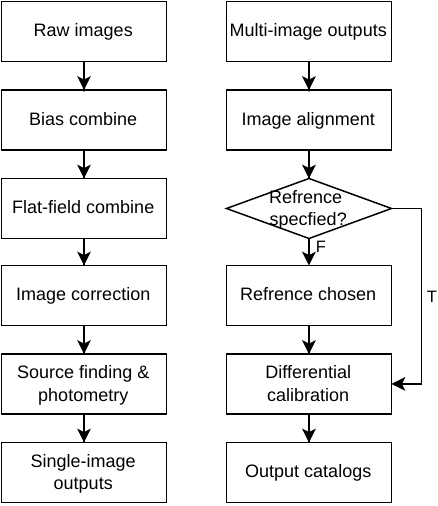} 
    \caption{General steps of the XZUT80LCP. Left: The single-image process; Right: The multi-image process}\label{flowchart}
\end{figure}

The XZUT-80LCP has a modular structure. It comprises modules for bias and flat-field combine, image correction, source finding and photometry, image alignment, reference star chosen, differential calibration. In addition, the pipeline includes a fast unified call module. The XZUT-80LCP flow chart, shown in Figure~\ref{flowchart}, is specifically controlled by the function parameter `step’. In the default configuration, the pipeline can automatically identify the observation data type. Subsequently, based on the pixel positions of the target, reference and check stars specified by the user, as well as other relevant parameters, the pipeline sequentially execute the photometric data processing steps. If the parameters of the reference and check stars are not provided, the pipeline will automatically execute the module for automatically selecting the reference star, and then select the appropriate reference and check stars. Once the reference star chosen module has been run, the pipeline outputs the pixel position and other information of the chosen reference stars. XZUT-80LCP supports single-image and multiimage process mode. Thanks to its modular design, users can selectively execute or replace specific modules.

\subsection{Program Advantages}

The XZUT-80LCP offers distinct advantages in the following areas:

\begin{itemize}
\item It is easy to call. When the pipeline is invoked, the entire steps from file recognition to light curve extraction can be completed automatically by specifying the function parameters on a one-off basis.  
\item It is highly modular. Program modules are coupled through input and output files. Users can replace or utilise specific modules, provided that the pipeline can still operate normally.
\item The pipeline is intuitive. After running each module, the program automatically generates the corresponding output file. It supports a variety of formats, enabling users to adjust the parameters efficiently and thereby optimise the pipeline’s processing results.
\item The reference stars are picked automatically. The program analyses photometric data to select the best reference stars for each band. These are usually bright stars with stable luminosity during the observation period.
\end{itemize}

\section{Program Modules}

\subsection{Image Correction}

After identifying the relevant files, the program preprocesses the raw data by performing bias correction, flat field correction, and image correction. Bias and flat field correction both employ median merging. Currently, most professional astronomical telescopes use CCD as detector and are equipped with cooling systems to reduce detector temperature. The dark current generated by the detector is low and generally does not require correction. Therefore, the dark correction step is not included in this pipeline. After preprocessing, the corrected image data are saved to the specified output directory with the suffix `\_bf.fits'. 

Due to the lack of unified specifications during the initial configuration phase of various telescopes, the metadata contained in FITS headers exhibits significant differences. Therefore, this program refers to the Xinglong Observatory Public FITS Header Standard (XPFHS) \citep{Lin2012} and reconstructs the FITS header. This will serve as the FITS file header template used by XZUT-80LCP when during image correction step. The FITS header template can be divided into the following components: basic information about the FITS file and station; the telescope and terminal instruments; the observed target; and the observation environment at the time.

\subsection{Source Finding and Photometry}

\del{SExtractor is software that creates a catalogue of objects from an astronomical image. It performs reasonably well on moderately crowded star fields and has been designed to be fast and robust. As SExtractor is comprehensive astronomical photometry software, highly accurate photometry results can be obtained simply by configuring the parameters. XZUT-80LCP considers this software to be the core functional component of the Star Finding and Photometry module. The program uses two SExtractor photometry modes: aperture and Kron. When Kron photometry is used, the software calculates the geometric parameters of the stars in the photometry results. }

\add{SExtractor is a comprehensive photometric software. Simply configuring the parameters allows it to automatically complete steps such as background subtraction, source finding and photometry. The XZUT-80LCP considers SExtractor to be the core component of the Source Finding and Photometry module, using two SExtractor photometry modes: circular aperture and Kron.}

\del{All relevant configurations have been preset. When invoking the program, only the `aper' parameter needs to be specified. This parameter supports the input of a set of photometry circul aperture values in pixels and can also be a negative value (a negative value indicates a multiple of the image's average full width at half-maximum, typically ranging from 2 to 3). For a more in-depth analysis of this photometry step, users can edit the default.sex file, which is located in the XZUT-80LCP package directory, and then place the modified file in the current working directory to use it directly. The photometry results will be saved with the `\_cat.fits' suffix. The `\_cat.fits' file mainly contains information such as star flux, flux error, geometric parameters and pixel position. }

\add{When using the source finding and photometry function, only the `aper' parameter needs to be entered. This parameter specifies the diameter used for circular aperture photometry. It can be set to a range of pixel values and can also be negative. Due to changes in weather conditions and other factors during observation, the FWHM will change. Using the same aperture size throughout the entire image will result in increased photometric error. Based on this, the function first calls SExtractor to measure the typical FWHM of the brightest stars in the image. It then calls SExtractor again, using a multiple of this FWHM to measure the star flux in the image. At this stage, a negative value represents a multiple of the FWHM, typically set to 2-3. Regardless of whether the `aper' parameter is specified, Kron photometry will be carried out and fields such as MAG\_AUTO will be output. The accuracy of Kron photometry is similar to that obtained using the optimal aperture for aperture photometry. The results of function execution will be saved with the catalogue, which contains fields such as pixel position, flux, flux error and geometric parameters.}

\subsection{Image Alignment}

The XZUT-80 features an equatorial mounting. Such telescopes only produce translations between different images. Based on this feature, \del{the program} \add{this work} uses the qmatch package for image alignment \add{and star matching steps;} the relevant algorithm is introduced in \citet{Zhengetal2024}. \del{The fundamental principle of this algorithm is as follows: Firstly, the average values of the rows and columns of the base image and other image are calculated respectively. Then, the quotient is calculated by dividing the average values of these two images under different offset. The smaller the quotient, the better offset value between these two images.} \add{The qmatch is faster than other softwares with a satisfied accuracy.}

\del{The package has been compiled into the program. The image alignment module can be invoked to perform image alignment while the star matching function has been integrated into the reference star selection and catalog module.} \add{The image alignment is achieved by invoking the image alignment module, and star matching has been integrated into the reference star selection and differential photometry modules.} When using the image alignment \del{module} \add{function}, only the `offset\_max\_dis' parameter needs to be input.\del{The range of this parameter is generally 0-0.25, meaning ensure the maximum offset does not exceed a quarter of the base image size.}\add{This parameter represents the maximum pixel offset between images. It is generally recommended as a quarter of the width/height of the images; otherwise, image alignment will fail.} After operating the image alignment module, an offset distribution map will be produced for each image relative to the base image. This can assist with monitoring the telescope's operational status. \del{For star matching function, users can set the `match\_max\_dis' parameter to define the max distance for star matching }
\add{After image alignment, further star matching requires input parameters `match\_max\_dis' to define the maximum pixel distance to match stars} (typically 1.5 to 3 times the FWHM).\add{The image alignment step spends about 10\% of the total processing time.}

\subsection{Reference Star Selection}

When performing differential calibration, reference and check stars must be provided \del{within the field of view}. The XZUT-80LCP assists users in manually specifying these stars and also provides a reference star selection module. The \del{general steps} \add{fundamental principle} in the reference star selection module are as follows: \del{First, a star with a high signal-to-noise ratio is selected from the base image to serve as a standard. The differences between the instrumental magnitudes of the selected star in the base image and in the other corrected images are then calculated. Next, the instrumental magnitudes of the corrected image are subtracted to align with the brightness level of the base image. Finally, a large magnitude data table is generated, with the table elements being the calibrated magnitude values. The results will then be calculated and analysed, including statistics on the standard deviation of the magnitude of stars during the entire observation period and the non-matching rate (the total proportion of stars on the base image that could not be successfully matched in other images). Ultimately, the ten most outstanding stars will be selected as the output results.} \add{Initially, a base image, typically the first in the sequence, is designated as the reference frame. Bright stars in the subsequent images are cross-matched with this base image, and the instrumental magnitude differences between each image and the base image are calculated, applying outlier clipping to ensure robustness. These magnitude differences are then used to calibrate the instrumental magnitudes of all images to the zero-point of the base image. A master table containing the calibrated magnitudes of all bright stars across the entire dataset is then generated. By computing the standard deviation of these magnitudes, stars exhibiting the lowest variability are identified and designated as the reference and check stars. As well as the standard deviation, other things are also considered when checking how stable these stars are.}

\del{Most of the variables involved in this module are adjustable to achieve better results. } \add{The `pick\_err\_max' parameter refers to the maximum photometric error and specifies the selection range for bright stars in the base image. It is usually set to between 0.05 and 0.01. } \del{The parameter `pick\_err\_max' specifies the selection of stars in the base image through SNR and is usually set to 0.05 or 0.01. }The standard deviation of the selected reference stars is specified by the parameter `pick\_ref\_std' and is typically less than 0.02.

\section{Program Application}

\subsection{Quick Program Invocation}



The XZUT-80LCP has been packaged with the code, configuration files, and a user manual, and integrated into the XZUT-80 observational system. XZUT-80LCP has also been published on GitHub at https://github.com/Polaris-XC1221/XZUT80LCP.

When invoking the pipeline, observation files must be named as follows: bias\_001.fit, flat\_V\_001.fit and object\_V\_001.fit. Underscores are recommended for delineating information such as the object name, band and sequence number. Please note that, given SExtractor's exclusive compatibility with Linux and macOS operating systems, the pipeline is subject to the same limitations.

%

\subsection{Program Output}

The Table~\ref{output files} enumerates all the output files that were produced by the program. The files are predominantly in the FITS format, which is extensively utilised in the field of astronomy, and the Pickle format, which the program can rapidly read and process. The program also provides corresponding text files, which are formatted as tables, enabling users to constantly adjust the module parameters to achieve better results. It is further possible for users to process the output files in accordance with their research. 

\begin{table*}
\caption{Output files of XZUT-80LCP.}\label{output files}
\begin{tabularx}{\textwidth}{@{}lXl@{}}
\toprule
Output file & Description & Format \\
\midrule
bias & Combined bias image  & Simple image FITS \\
ﬂat\_(band) & Combined flat images of all bands & Simple image FITS \\
(object)\_(band)\_(number)\_bf  & Corrected science images of all objects and bands & Simple image FITS \\
(object)\_(band)\_(number)\_cat & Catalogs of find and photometry step in configuration file & Binary table FITS \\
(object)\_(band)\_(number)\_se & Catalogs of find and photometry step in bright star standards & Binary table FITS \\
offset\_(object)\_(band) & X\&Y offsets between images and reference image, offsets movement map & Pickle, Text, PNG \\ 
pick\_(object)\_(band) & Chosen reference stars information, Candidate Reference Stars Certification Diagram & Binary table FITS, Pickle, Text, PNG \\
cata\_(object)\_(band) & General catalog of target, reference and check stars, Certification Diagram & Binary table FITS, Pickle, PNG \\ 
cali\_(object)\_(band) & Differential calibrated catalog of target and check stars calibrated by reference stars & Binary table FITS, Pickle, Text \\
lc\_(obj)\_(band) & Differential calibrated Light-curve graph & PNG \\
\bottomrule
\end{tabularx}
\end{table*}

\subsection{Program Usage Example}

This section provides an illustration of the program's application, utilising observational data obtained on 31 October 2024 from the exoplanet candidate TOI3646.01. The data presented was collected using the XZUT-80 and its configured 2K camera. A total of 400 images were captured during the period without the use of a filter. Furthermore, ten bias images and five flat-field images in the none band were obtained. During the observation period, the weather conditions were optimal, with clear skies and no clouds. On the night in question, the moon was in its waning phase. Reference stars and check stars have also been specified. The primary star of this object has a magnitude of $m_{v} = 14.164 \pm 0.103$ in the V band.


The XZUT-80LCP processes the images in two modes. The call details are shown in Listing~\ref{example code1} and Listing~\ref{example code2}, and the total runtime is approximately 20 and 30 min, respectively. Invoking the function obtains the following parameters: The paths of the raw data (raw\_dir) and output directory (red\_dir); the object’s band (band), name (object) and coordinates (alt\_coord), and the positions of the target, reference and check stars within the base image (starxy); and the index numbers of the target, reference and check stars in the list (ind\_tgt, ind\_ref, ind\_chk). The function uses the `aper’ parameter to specify a range of aperture sizes and the ‘base\_img’ parameter to designate the sixth science image data as the reference image.

\begin{lstlisting}[caption={\del{Taking the exoplanet TOI3646.01 as an example, here is how to call the program.} \add{The calling code of XZUT-80LCP when manually specifying the target, reference and check star.}}, label={example code1}]
import xzut80lcp

xzut80lcp.run(
    raw_dir="/mnt/f/raw/20241031",
    red_dir="/mnt/f/red/20241031",
    band="N",
    object="TOI3646",
    alt_coord=("00:58:41.43","+57:35:24.23"),
    aper=[17,18,19,20,21, 
          -2,-2.5,-3],
    base_img=6,
    starxy=[(1030,1054),(985,865),(1229,667)],
    ind_tgt=0,ind_ref=1,ind_chk=2,
)
\end{lstlisting}

\begin{lstlisting}[caption={\add{The calling code of XZUT-80LCP when the Reference Star Selection module is running}}, label={example code2}]
import xzut80lcp

xzut80lcp.run(
    raw_dir="/mnt/f/raw/20241031",
    red_dir="/mnt/f/red/20241031",
    band="N",
    object="TOI3646",
    alt_coord=("00:58:41.43","+57:35:24.23"),
    aper=[17,18,19,20,21, 
          -2,-2.5,-3],
    base_img=6,
    starxy=[(1030,1054)],
)
\end{lstlisting}

\subsection{Output Result Description}

\begin{figure}
	\centering
	\includegraphics[width=0.5\textwidth]{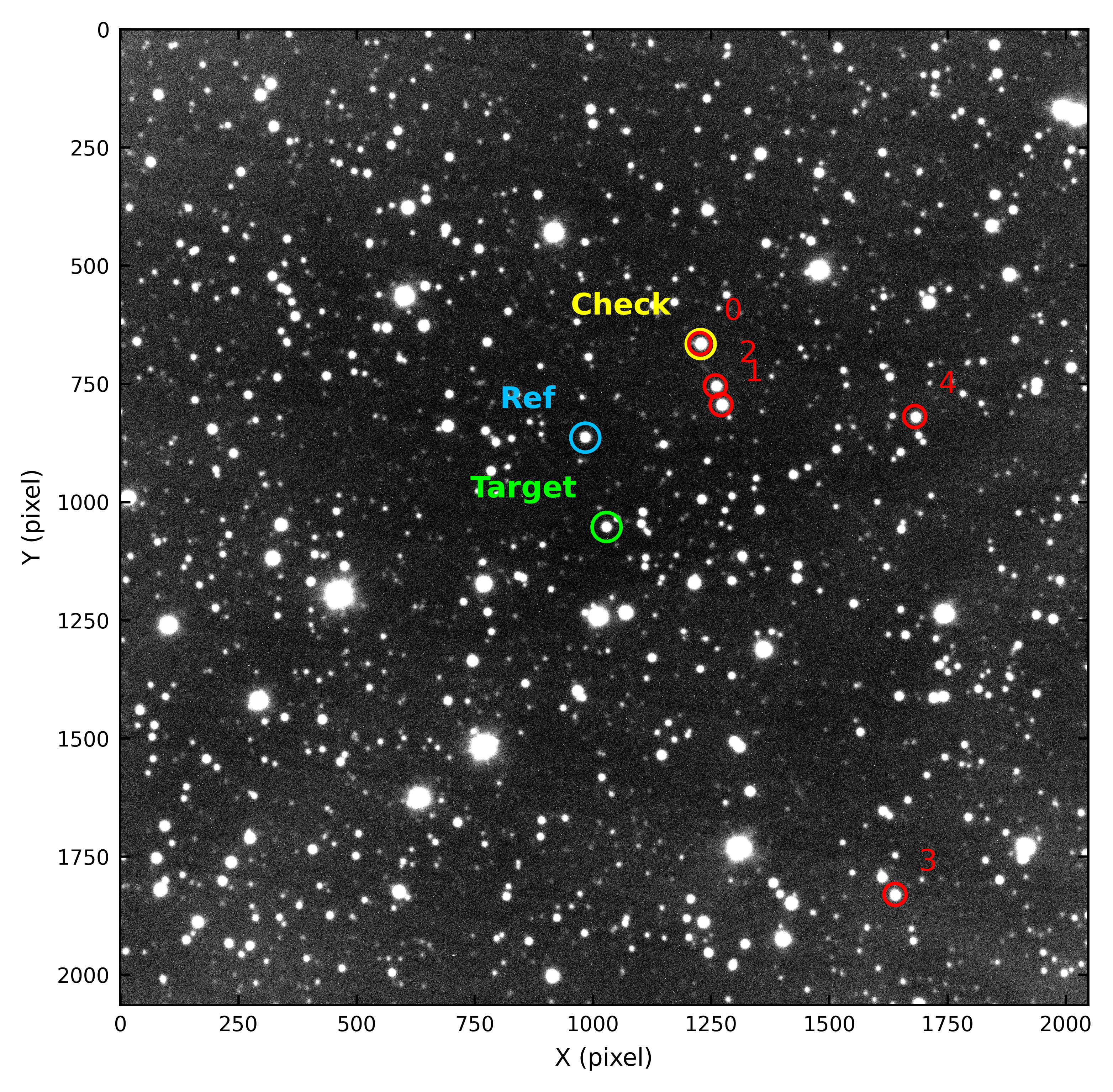}	
	\caption{\del{Finding chart of the target, reference and check stars generated by the program. In this diagram, star 0 is the target star, star 1 is the reference star, and star 2 is the check star.}\add{On October 31, 2024, the N-band image of the exoplanet TOI3646 was observed using XZUT-80. The green marked is the target star. The blue and yellow are the manually specified reference and check stars. The red stars numbered 0 and 1 to 4 are the automatically selected check and reference stars.}}\label{Confirmation diagram}
\end{figure}

\begin{figure}
	\centering
	\includegraphics[width=0.5\textwidth]{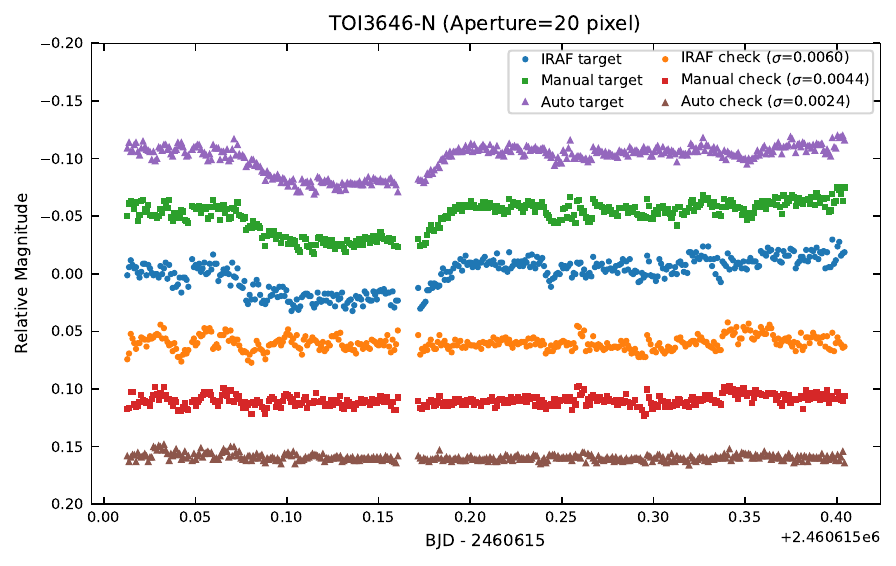}	
	\caption{A diagram of the light curve. \del{It includes the results obtained using XZUT-80LCP and IRAF. The curves from above are checker by this work, checker by IRAF, target by this work, and target by IRAF, respectively. }\add{In the above curves, the triangles represent the target and check by the XZUT-80LCP's automatic reference star selection mode, the squares represent the target and check by the XZUT-80LCP's manual specified reference star mode,  the circles represent the target and check by the IRAF. The selection of stars is consistent with that shown in Figure~\ref{Confirmation diagram}}}\label{lightcurve}
\end{figure}

\begin{figure}
	\centering
    \includegraphics[width=0.5\textwidth]{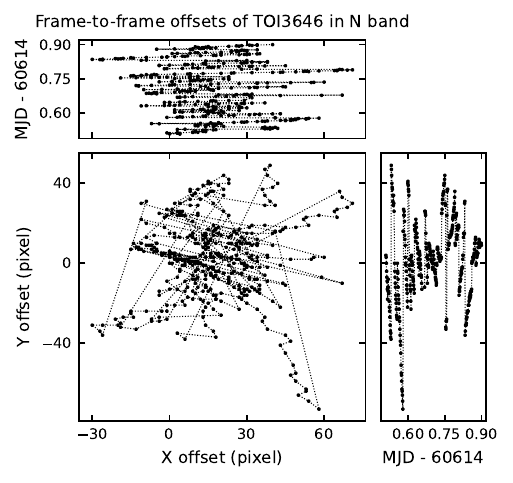}
	\caption{This shows the offset of the telescope's pointing during the observation. \add{The dashed lines between points represent the telescope’s offset between two consecutive exposures.}}\label{offset_map}
\end{figure}

For instance, in the case of the data of TOI3646.01, an exoplanet candidate, in the Null band, the primary outputs of the program are as follows: Figure~\ref{Confirmation diagram} , the finding-chart of the target star, the reference stars and the checker star. Figure~\ref{lightcurve} illustrates the light-curves of the target and check stars using differential photometry. With an aperture size of 20 pixels, the standard deviation of the check star in the N band is 0.0044. The result is of the same order of magnitude as the dispersion in the light curve caused by the lower photometric accuracy of ground-based telescopes. Furthermore, the program produced an offset map, as demonstrated in Figure~\ref{offset_map}, which shows the tracing of the telescope. 



\subsection{Compared with the result from IRAF}

As comparation, data were also processed using IRAF. Photometry was performed using the noao.digiphot.apphot.phot task. A fixed aperture of 20 pixels was selected for photometry, with the sky background ring ranging from 35 to 45 pixels.

\section{Conclusions}

The XZUT-80LCP is an automated pipeline for processing photometric data and extracting light curves, specifically for variable stars. It has been integrated into the XZUT-80 system to address deficiencies in data processing. This paper outlines the program's usage methods, workflow, core modules, and processing results. Additionally, it introduces several excellent softwares utilized in the pipeline, such as SExtractor for star finding and photometry, and QMatch for image alignment. We processed photometric data of the exoplanet candidate TOI3646.01, which was observed on October 31, 2024. The results show that the differential photometric outcomes of this program are comparable in magnitude to those produced by established photometric processing pipelines for 1.26m Infrared Telescope \citep{Zhongetal2018} and QLCP \citep{ZhengJiang2023}, while also demonstrating greater precision and efficiency than the commonly used IRAF in the astronomical community. 

Although the functions of this pipeline are not as extensive as those of larger-scale data processing systems, it possesses notable qualities such as being lightweight, automated, modular, and is highly suited for the XZUT-80. Should the XZUT-80 camera be replaced in the future, only minor adjustments to necessary parameters would be required for the program to continue functioning properly. However, it is important to acknowledge that this pipeline currently lacks astrometry processing capabilities, which presents certain limitations. Future efforts will be directed towards addressing this shortcoming.

\section*{Acknowledgements}

This work was supported by the Strategic Priority Research Program of the Chinese Academy of Sciences [grant number XDB0550103]. We acknowledge the support of the staff of the Xinglong Observatory, NAOC. This work was partially supported by National Astronomical Observatories, Chinese Academy of Sciences.

\printcredits

\bibliographystyle{cas-model2-names}

\bibliography{reference}



\end{document}